\documentclass[fleqn,usenatbib]{mnras}

\usepackage{newtxtext,newtxmath}

\usepackage[T1]{fontenc}

\DeclareRobustCommand{\VAN}[3]{#2}
\let\VANthebibliography\thebibliography
\def\thebibliography{\DeclareRobustCommand{\VAN}[3]{##3}\VANthebibliography}

\usepackage{graphicx}	
\usepackage{amsmath}	
\usepackage{caption}
\usepackage{subcaption}

\newcommand{\msun}{${\rm M}_{\odot}$\,}

\title[Photometric masses of CSS131106]{Photometric masses for long period CVs: the case study of CSS131106}

\author[M. Das et al.]{
M. Das$^{1}$, 
S. P. Littlefair$^{2}$\thanks{E-mail: s.littlefair@sheffield.ac.uk (SPL)},
S. G. Parsons$^{2}$,
V. S. Dhillon$^{2, 3}$,
M. J. Dyer$^{2, 4}$,
A. J. Brown$^{5}$,
J. A. Garbutt$^{2}$,\newauthor
M. J. Green$^{6, 7}$,
D. Jarvis$^{2}$,
M. R. Kennedy$^{8}$,
P. Kerry$^{2}$,
E. Pike$^{2}$,
D. I. Sahman$^{2}$,
Amalie Yates$^{2}$,
J. McCormac$^{9}$, \newauthor
N. Castro Segura$^{9}$,
J. Munday$^{9}$,
I. Pelisoli$^{9}$
\\
$^{1}$Department of Physics and Earth Science, University of Ferrara, Via Saragat 1, I-44122 Ferrara, Italy\\
$^{2}$Astrophysics Research Cluster, School of Mathematical and Physical Sciences, University of Sheffield, Sheffield, S3 7RH, UK\\
$^{3}$Instituto de Astrof\'isica de Canarias, E-38205 La Laguna, Tenerife, Spain\\
$^{4}$Research Software Engineering, University of Sheffield, Sheffield, S1 4DP, UK\\
$^{5}$Hamburger Sternwarte, University of Hamburg, Gojenbergsweg 112, 21029 Hamburg, Germany\\
$^{6}$Homer L. Dodge Department of Physics and Astronomy, University of Oklahoma, 440 W. Brooks Street, Norman, OK 73019, USA\\
$^{7}$JILA, University of Colorado and National Institute of Standards and Technology, 440 UCB, Boulder, CO 80309-0440, USA\\
$^{8}$School of Physics, University College Cork, Cork, T12 K8AF, Ireland\\
$^{9}$Department of Physics, University of Warwick, Gibbet Hill Road, Coventry, CV4 7AL, UK
}

\date{Accepted XXX. Received YYY; in original form ZZZ}

\pubyear{\the\year{}}

\begin{document}
\label{firstpage}
\pagerange{\pageref{firstpage}--\pageref{lastpage}}
\maketitle

\begin{abstract}
We present high-speed photometry of the eclipsing cataclysmic variable CSS131106 J052412+004148. We determine the system parameters by modelling the eclipse lightcurve using the photometric eclipse method, in which the mass ratio is determined from the relative timings of the white dwarf and bright spot eclipses. Despite the blended white dwarf and bright spot ingress, typical of longer period cataclysmic variables, we perform simulations that show we are able to reliably constrain the component masses. We find a mass ratio of $q = 0.81 \pm 0.06$ and inclination $i = 78.5 \pm 0.7$ degrees. The white dwarf and donor masses were found to be $M_{w} = 0.72 \pm 0.04$~\msun{} and $M_{d} = 0.58 \pm 0.06$~\msun{} respectively. The white dwarf temperature was estimated to be $T_{\rm eff} = 18~500 \pm 2~000$~K, implying a moderate accretion rate of $\dot{M} = 3 \pm 1 \times 10^{-10}$~\msun{}~yr$^{-1}$. The donor in CSS131106~J052412+004148 joins two other long-period cataclysmic variables (IP~Peg and HS~0220+0031) in being unusually small for its mass, even when compared to detached M-dwarfs. The donors in all three systems are also unusually cool for their mass. We discuss possible explanations for the small radii and cool temperatures of the donors in these systems, but find no viable explanation for their properties.
\end{abstract}




\section{Introduction}
Precise measures of the masses and radii of donors and white dwarfs in cataclysmic variables (CVs) are difficult to obtain, because the accretion disc often outshines both the donor and the white dwarf. As a result, CVs are almost never observed as double-lined eclipsing binaries. Masses can be derived spectroscopically if the donor is visible, by assuming that the motion of the disc's emission lines trace the orbital motion \cite[e.g][]{marsh87}, but this assumption is not always true, and spectroscopic mass determinations also require an uncertain correction between the centre of mass and the centre of light of the irradiated donor (\citealt{parsons10}).

The majority of precise component masses in CVs have been measured by modelling the shape of the primary eclipse, hereafter called the {\em photometric eclipse method} (e.g. \citealt{wood89a, feline04a, littlefair06, savoury11, mcallister2019}). The photometric eclipse method relies on being able to accurately measure the ingress and egress phases of both the white dwarf and the bright spot. We recap the principles by which the method works here. The white dwarf ingress and egress phases allow us to measure the width of the white dwarf eclipse $\Delta \phi_{1/2}$, which is a function only of the mass ratio $q$ and inclination $i$. We make the assumption that the gas stream from the donor follows a free-fall path. In that case, assuming the bright spot is small in extent and eclipsed rapidly, the ingress and egress phases of the bright spot ($\phi_{BS}^i, \phi_{BS}^e$) depend only upon $q, i$ and the accretion disc radius $R_d$. We therefore have three unknowns $\{q, i, R_d\}$ and three constraints $\{\Delta \phi_{1/2}, \phi_{BS}^i, \phi_{BS}^e\}$, which allows us to find the mass ratio $q$. The duration of white dwarf ingress/egress allows us to find the white dwarf radius and (assuming a mass-radius relationship) the white dwarf mass, $M_w$. The other system parameters follow from $M_w$, $q$ and Kepler's third law.

In reality, the bright spot eclipse is not sudden, because the bright spot is a complex structure with a non-negligible size. In its modern application, the photometric eclipse method uses a parameterised model of the bright spot \citep{copperwheat10} to relax the assumption of a compact bright spot. It is assumed that the brightest part of the spot arises where the gas stream hits the disc. The model of the bright spot is not physical, but designed to be able to recreate a wide variety of bright spot eclipse lightcurves. Although rarely mentioned, the finite extent of the bright spot and any failure of the parametric model to replicate the true bright spot eclipse lightcurve are perhaps the most important systematic error of the photometric eclipse method. However, masses measured using the eclipse lightcurves have been independently confirmed in systems where it is also possible to measure radial velocities (e.g. \citealt{tulloch09, copperwheat2012, savoury12, littlefair24}), so the assumptions that underpin the photometric eclipse method seem to be valid in the systems to which the method has been applied so far.

The application of the photometric eclipse method to CVs above the period gap is hampered by two factors. Firstly, the population is dominated by nova-like variables, where the white dwarf and bright spot eclipses are typically not visible due to the brightness of the disc itself (see \citealt{dhillon13}, for example). However, even in dwarf novae, long period systems tend to have lightcurves in which the bright spot ingress is extended in duration, and where white dwarf ingress occurs at the same time as the bright spot ingress, leading to these two features being blended. This makes parameter determination challenging as the blending of white dwarf and bright spot ingress complicates the measurement of the white dwarf eclipse width $\Delta \phi_{1/2}$.

In principle it is still possible to use lightcurves with blended ingresses to measure masses - after all the fact that the two features are blended gives some information about the timing of white dwarf ingress, and the white dwarf ingress will subtly change the shape of the bright-spot ingress feature, so the timing of white dwarf and bright spot ingress is encoded in the lightcurve data, allowing accurate measurement of $\Delta \phi_{1/2}$. However, it is reasonable to worry that a mismatch between the model bright spot ingress and the true bright spot ingress could have an effect on the derived parameters for systems with blended ingresses. In this paper, we present a case study of applying the photometric eclipse method to a long period CV with blended ingresses, to assess the viability of the technique.

CSS131106 J052412+004148 (hereafter CSS131106) was discovered through multiple outbursts by the Catalina Real-Time Transient Survey (CRTS, \citealt{drake09}). Its eclipsing nature was identified by \cite{hardy17}. High-speed photometry in that paper showed that the ingresses of the white dwarf and bright spot were blended, making it an excellent test case to see if photometric masses can still be measured in this scenario.

\section{Observations and Data Reduction}
\label{sec:obs}
Observations of CSS131106 were taken with the triple-band fast camera ULTRACAM \citep{dhillon07} between the dates of 
November 2016 and January 2018. ULTRACAM is mounted on the 3.5-m New Technology Telescope (NTT) located at La Silla, Chile. The ULTRACAM data were taken in filter sets similar to the Sloan Digital Sky Survey (SDSS) $u'$, $g'$ and $r'$ passbands. Observations in 2016 were taken in standard SDSS filters, but observations from 2017 onwards were taken in SDSS-like filters designed for enhanced throughput, which we refer to as Super-SDSS ($u_s, g_s, r_s$) - see \cite{dhillon21} for details. A full table of observations is presented in Table~\ref{table:obs}.

\begin{table*}
\begin{center}
\caption[]{Journal of observations.  To improve the signal-to-noise ratio the blue CCD was read out less frequently than the green and red CCDs. The column Nblue shows the number of individual exposures before the blue CCD was read out. The dead-time between exposures was 0.025~s for all observations. The GPS timestamps on each data point are accurate to 50 $\mu$s.}
\begin{tabular}{crrccccccccc}
\hline
Date & Start Phase & End Phase  & Filters & Exposure time (s) & Nblue &
Data points & Cycle Number & Seeing (arcsec) & Airmass \\
\hline
2016-11-09 & -0.163 & 0.183 & $u'g'r'$ & 3.08 & 1 & 1687 & -2020 & 0.7--0.8 & 1.16--1.27 \\
2018-01-17 & -0.116 & 0.142 & $u_{s}g_{s}r_{s}$ & 2.96 & 3 & 1304 & 464 & 0.8--1.0 & 1.15--1.20 \\
2018-01-25 & -0.133 & 0.151 & $u_{s}g_{s}r_{s}$ & 3.08 & 1 & 1385 & 510 & 0.9--1.1 & 1.22--1.45 \\
2018-01-27 & -0.214 & 0.157 & $u_{s}g_{s}r_{s}$ & 3.08 & 4 & 1811 & 521 & 0.9--1.0 & 1.15--1.18 \\
2018-01-28 & -0.165 & 0.131 & $u_{s}g_{s}r_{s}$ & 3.08 & 1 & 1444 & 527 & 0.9--1.1 & 1.17--1.30 \\
\hline
\hline
\end{tabular}
\label{table:obs}
\end{center}
\end{table*}

Data reduction was carried out using the HiPERCAM pipeline software\footnote{https://github.com/HiPERCAM/hipercam}. Optimal photometry \citep{naylor98} was used with a variable aperture radii of 1.7 times the full width at half maximum (FWHM). A nearby comparison star was used to correct for transparency variations. For flux calibration, the SDSS magnitudes of the comparison stars were converted in the natural photometric system for the ULTRACAM filters using the colour terms in \cite{wild22}.

\subsection{Orbital Ephemeris}
The middle of the white dwarf eclipse is taken to be the fiducial point for the ephemeris. Mid-eclipse times were determined by averaging the time of white dwarf ingress and egress, as determined by locating the minima and maxima of a smoothed lightcurve derivative. Mid eclipse times were corrected to the Solar System Barycentre using {\sc astropy} \citep{astropy}. Mid-eclipse times are presented in table~\ref{tab:eclipsetimes}. A linear ephemeris was fitted to the eclipse times to give an ephemeris of:
\begin{equation}
\label{eq:ephem}
T({\rm BMJD}) = 58055.07489(2) +  0.17466676(2) E.
\end{equation}

\begin{table}
\centering
\caption{Mid-eclipse times of CSS131106.}
\label{tab:eclipsetimes}
\begin{tabular}{ccc}
\hline
Cycle Number & BMJD(TDB) & Uncertainty (days) \\
\hline
-2020  & 57702.24803 & 0.00005 \\
464    & 58136.12023 & 0.00005 \\
510    & 58144.15499 & 0.00005 \\
521    & 58146.07624 & 0.00005 \\
527    & 58147.12428 & 0.00005 \\
\hline
\end{tabular}
\end{table}

\begin{figure}
\centering
\includegraphics[width=\linewidth]{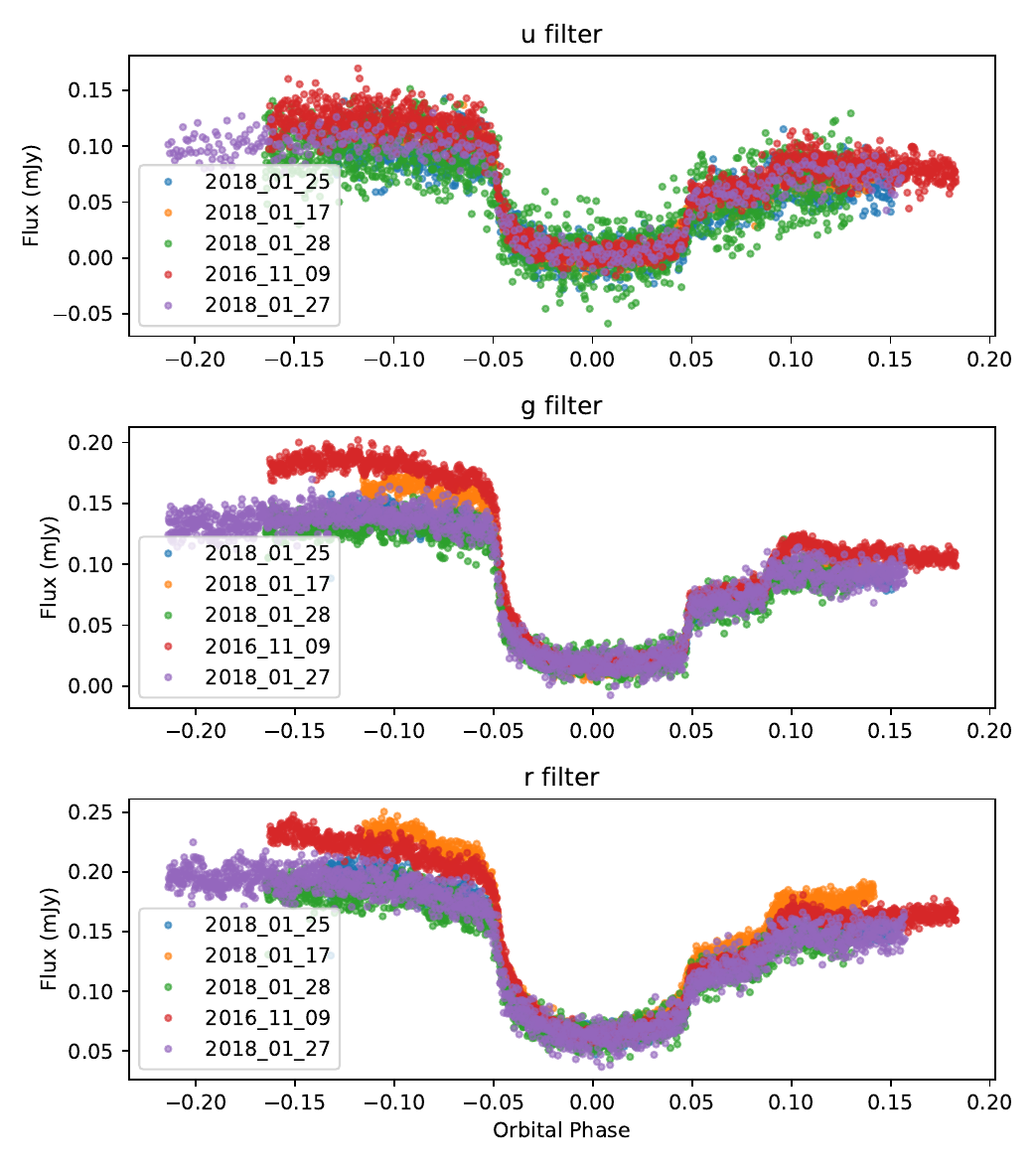}
\caption{Flux calibrated ULTRACAM lightcurves of CSS131106 in (top-to bottom) $u'$, $g'$ and $r'$ bands.}
\label{fig:all_lightcurves}
\end{figure}

The ULTRACAM lightcurves were phase folded using the ephemeris above and are shown in Figure~\ref{fig:all_lightcurves}. 

\section{Results}
\label{sec:results}

\subsection{Modelling eclipses with blended ingress features}
\label{subsec:sims}
As stated in the introduction, whilst it is theoretically possible to model eclipse lightcurves in which the white dwarf and bright spot ingresses are blended, we should be concerned that systematic errors in the bright spot model may skew the modelled mass ratio $q$ and white dwarf eclipse width $\Delta \phi_{1/2}$. Here we present simulations that show that this is not the case, and that accurate estimates of the mass ratio can be derived from these systems.

\begin{figure*}
\centering
\begin{subfigure}[b]{.99\linewidth}
\includegraphics[width=\linewidth, trim=.0cm .0cm .0cm 0.8cm, clip]{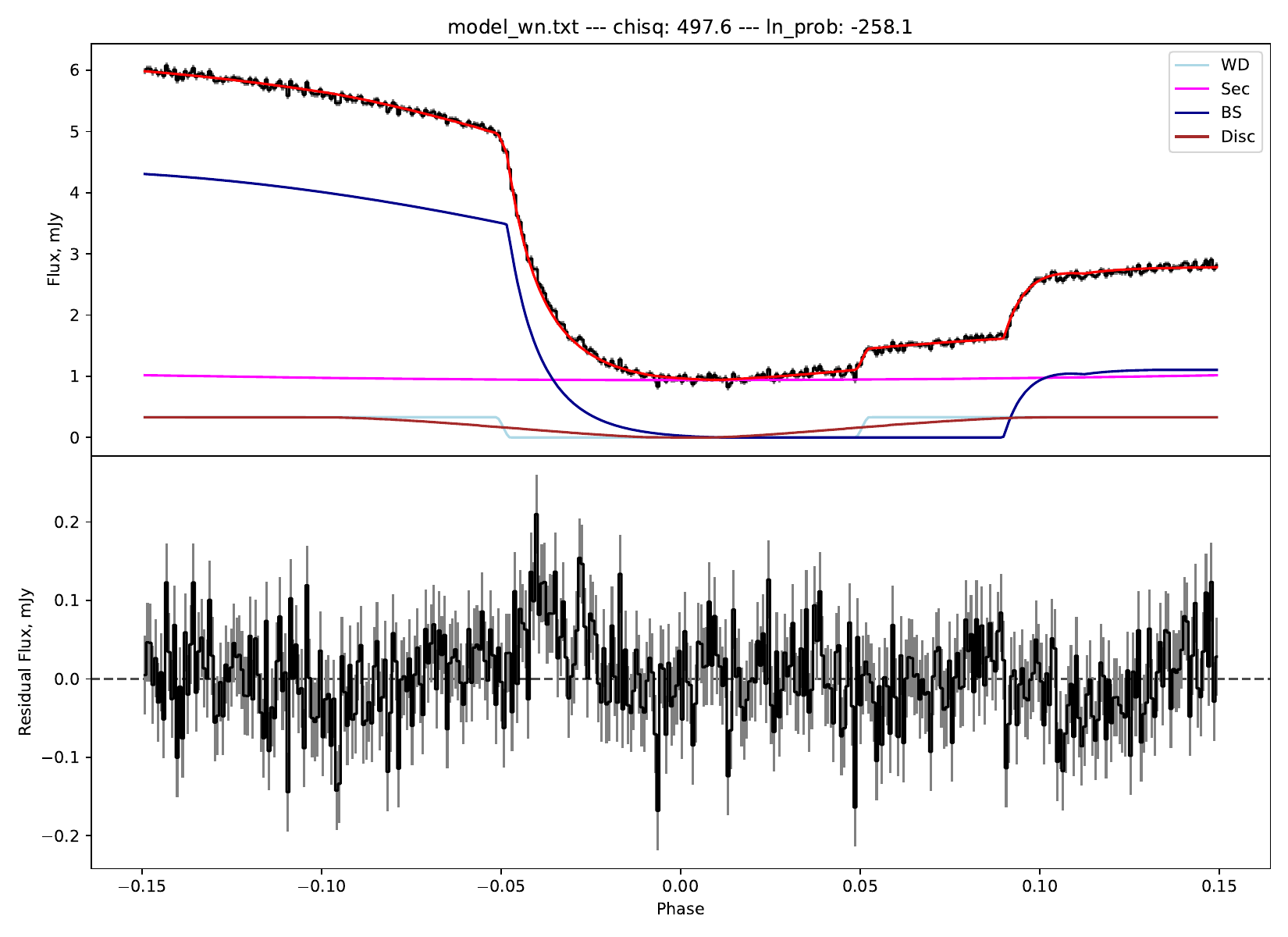}
\end{subfigure}
\caption{Model fit to simulated eclipse lightcurve with a blended white dwarf and bright spot ingress.}
\label{fig:symfit}
\end{figure*}

Using the binary modelling code {\sc lcurve} \citep{copperwheat10} one of the authors generated an eclipse lightcurve for a CV in which the bright spot and white dwarf ingresses are blended, but kept the model parameters blinded from the remaining authors. Gaussian noise was added to the lightcurve at typical levels seen in CV eclipse lightcurves and the eclipse was modelled using the {\sc lfit} code, as described in \cite{mcallister2017}. {\sc lfit} is a simplified version of {\sc lcurve}, which implements the same bright spot model. The resulting model fit is shown in figure~\ref{fig:symfit}. 

The simulated lightcurve has a very strong bright spot and relatively weak white dwarf egress, so the simulation maximises the potential for any errors in the bright spot model to affect system parameters such as the mass ratio and the inclination. The mass ratio and inclination for the simulated lightcurve are chosen from the relatively narrow range of parameters which produce a blended bright spot ingress. The inclination is potentially very difficult to constrain, as this affects both the ingress/egress of the bright spot, but also the ingress/egress of the white dwarf - and the effect of the small white dwarf ingress on the bright spot ingress is quite minor.

\begin{figure}
\centering
\begin{subfigure}[b]{.99\linewidth}
\includegraphics[width=\linewidth]{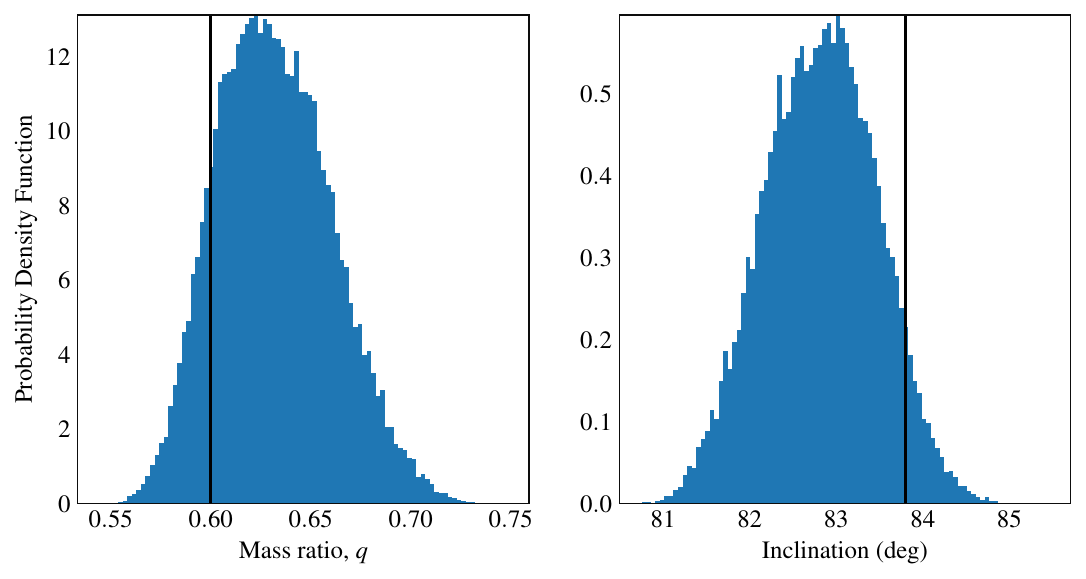}
\end{subfigure}
\caption{Posterior distributions of mass ratio and inclination compared to true values for the simulated lightcurve. Histograms show posterior probability samples from the MCMC chains. The black vertical lines denote the true values of the parameters.}
\label{fig:symfit_results}
\end{figure}

In figure~\ref{fig:symfit_results}, we show a comparison between the retrieved model parameters and the true values used to generate the lightcurve. We focus on the two parameters that have the most impact on the resulting system parameters: the mass ratio, and the inclination. Despite some small systematic residuals during bright spot ingress seen in figure~\ref{fig:symfit}, the modelled parameters are in excellent agreement with the true values, giving us confidence that accurate system parameters can be retrieved from eclipses with blended ingresses.

\subsection{Modelling of CSS131106}
\label{subsec:ecl_model}
\begin{figure*}
\centering
\begin{subfigure}[b]{.58\linewidth}
\includegraphics[width=\linewidth]{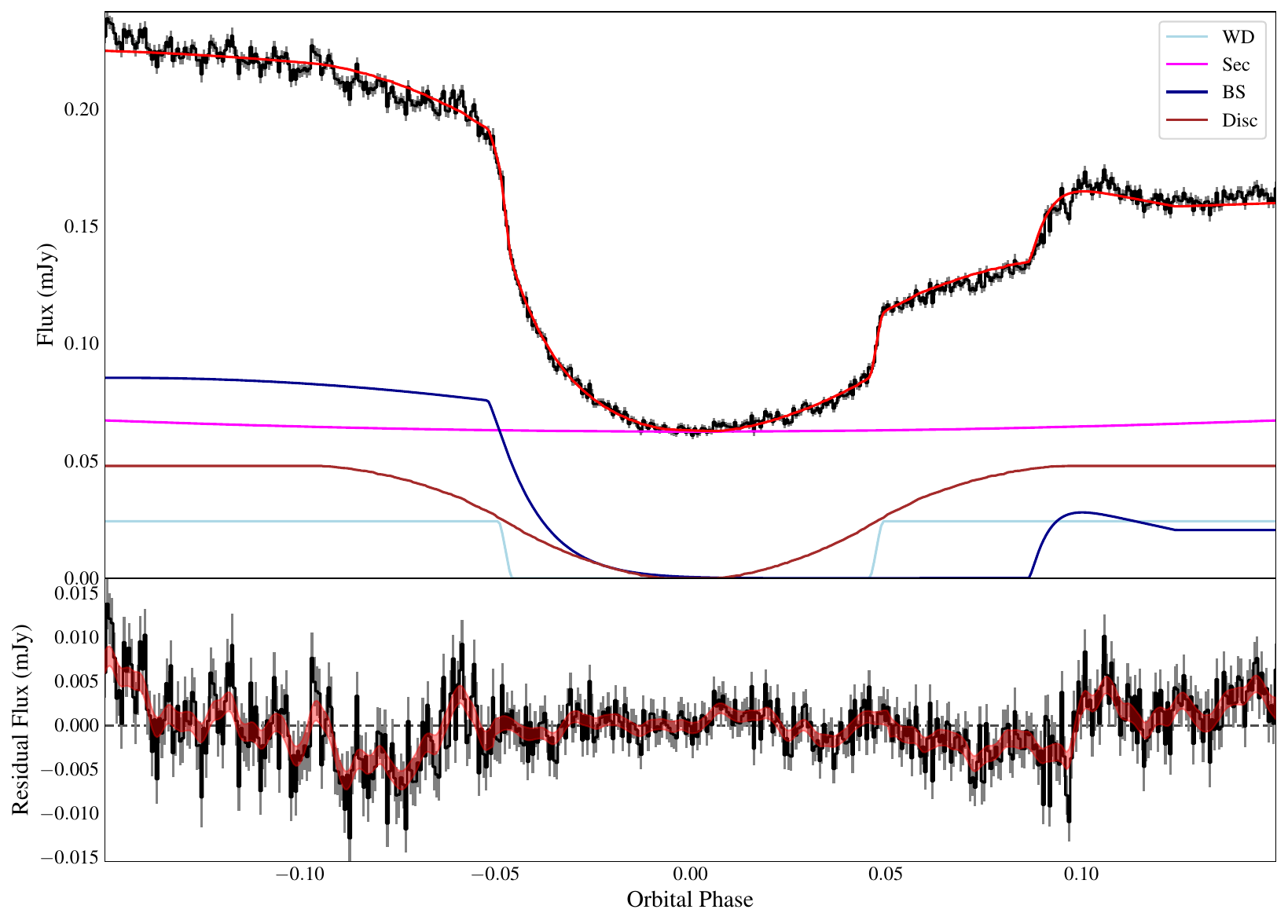}
\end{subfigure}
\begin{subfigure}[b]{.58\linewidth}
\includegraphics[width=\linewidth]{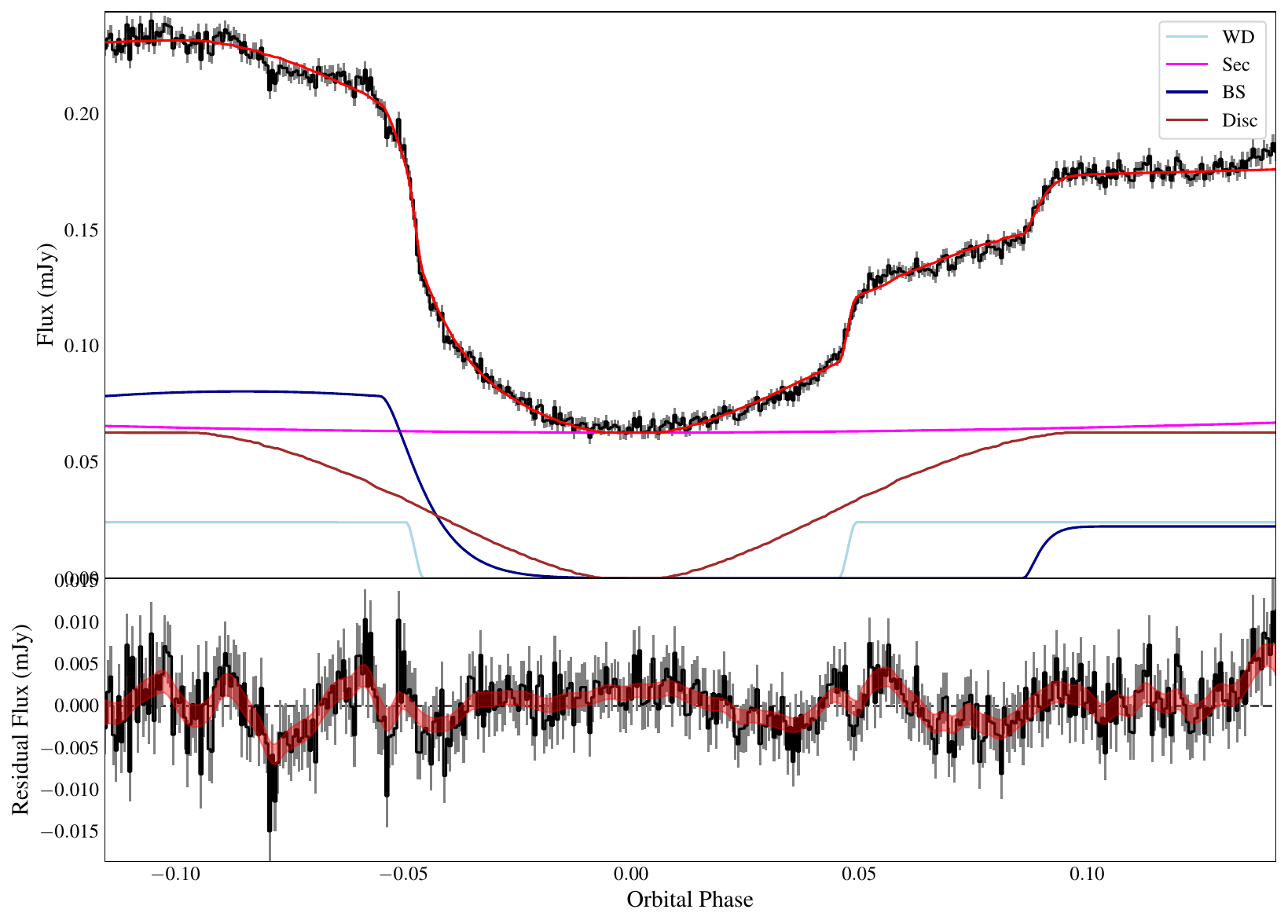}
\end{subfigure}
\begin{subfigure}[b]{.58\linewidth}
\includegraphics[width=\linewidth]{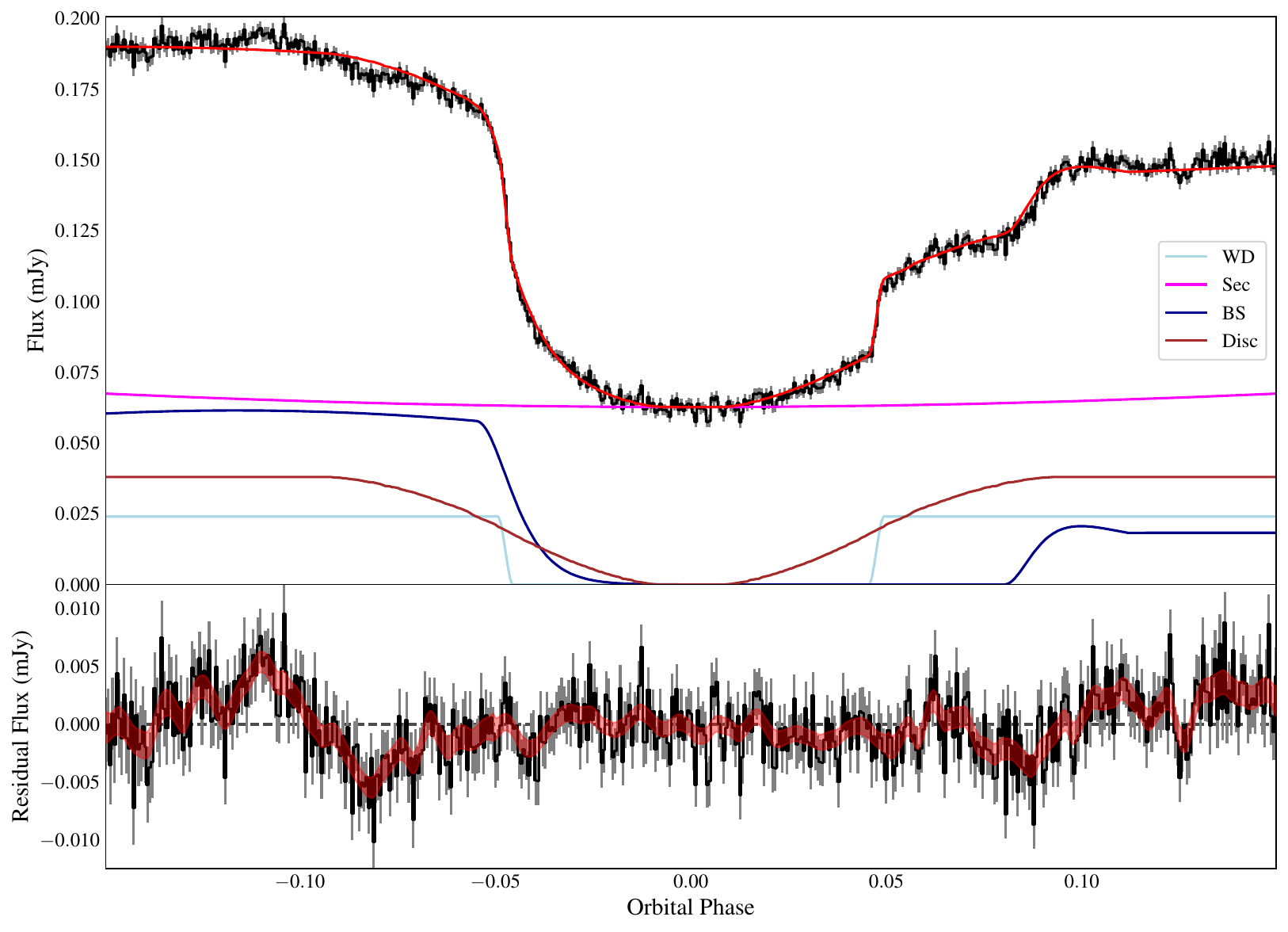}
\end{subfigure}
\caption{Phase folded and binned $r'$-band lightcurves together with model fits. From top to bottom, the panels show the 2016 data, the 2018-01-17 data, and the binned lightcurves of the remaining data. Data are shown as a black line with error bars. Model fits are shown in red. Also shown are the different components of the model: the white dwarf (cyan), bright spot (dark blue), accretion disc (dark red) and donor star (purple). The lower panels show the residuals of the fit (black line with errors) and the 1$\sigma$ posterior of the Gaussian process model to the residuals (red band).}
\label{fig:rmodelfits}
\end{figure*}

\begin{figure*}
\centering
\begin{subfigure}[b]{.58\linewidth}
\includegraphics[width=\linewidth]{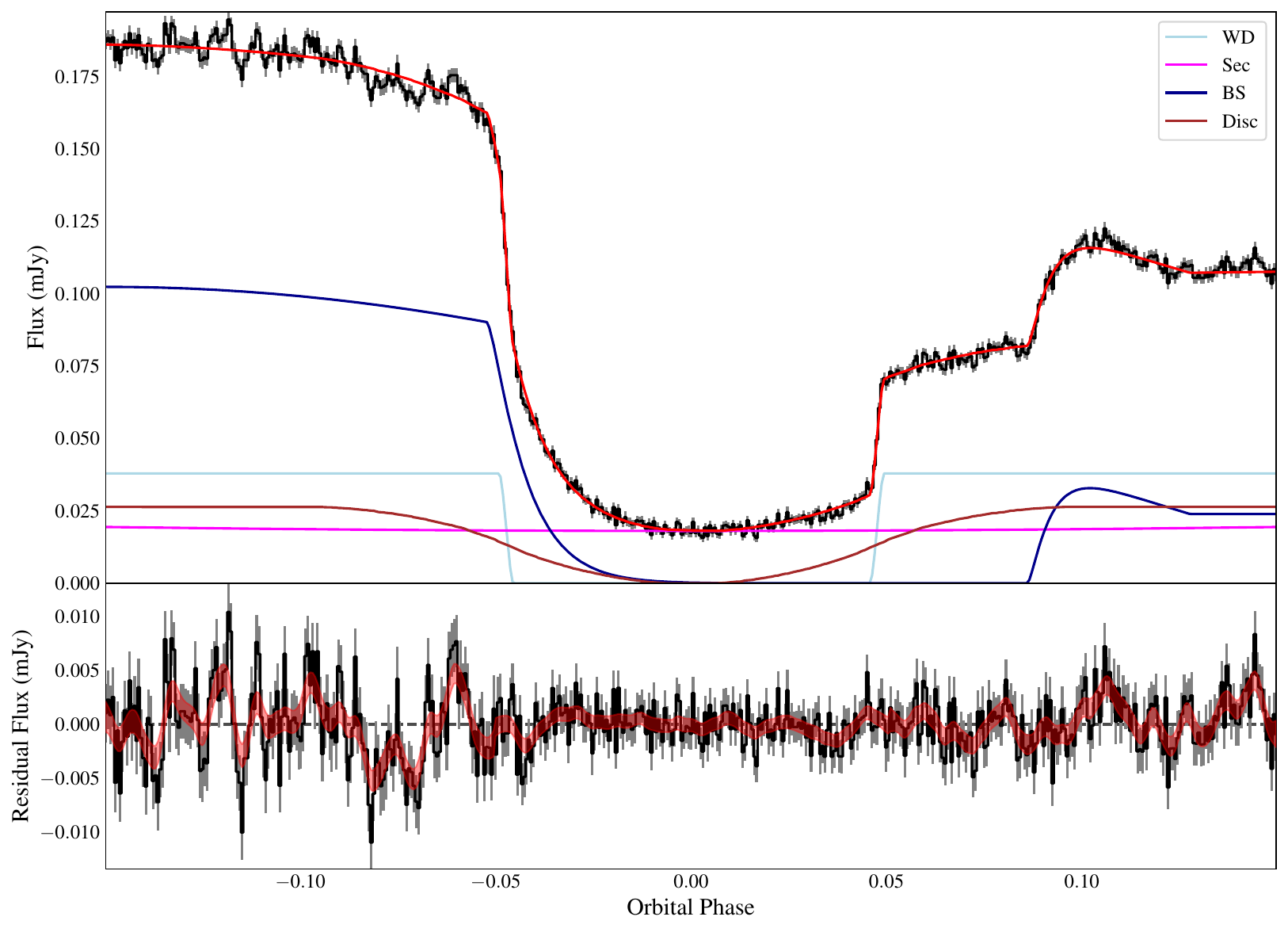}
\end{subfigure}
\begin{subfigure}[b]{.58\linewidth}
\includegraphics[width=\linewidth]{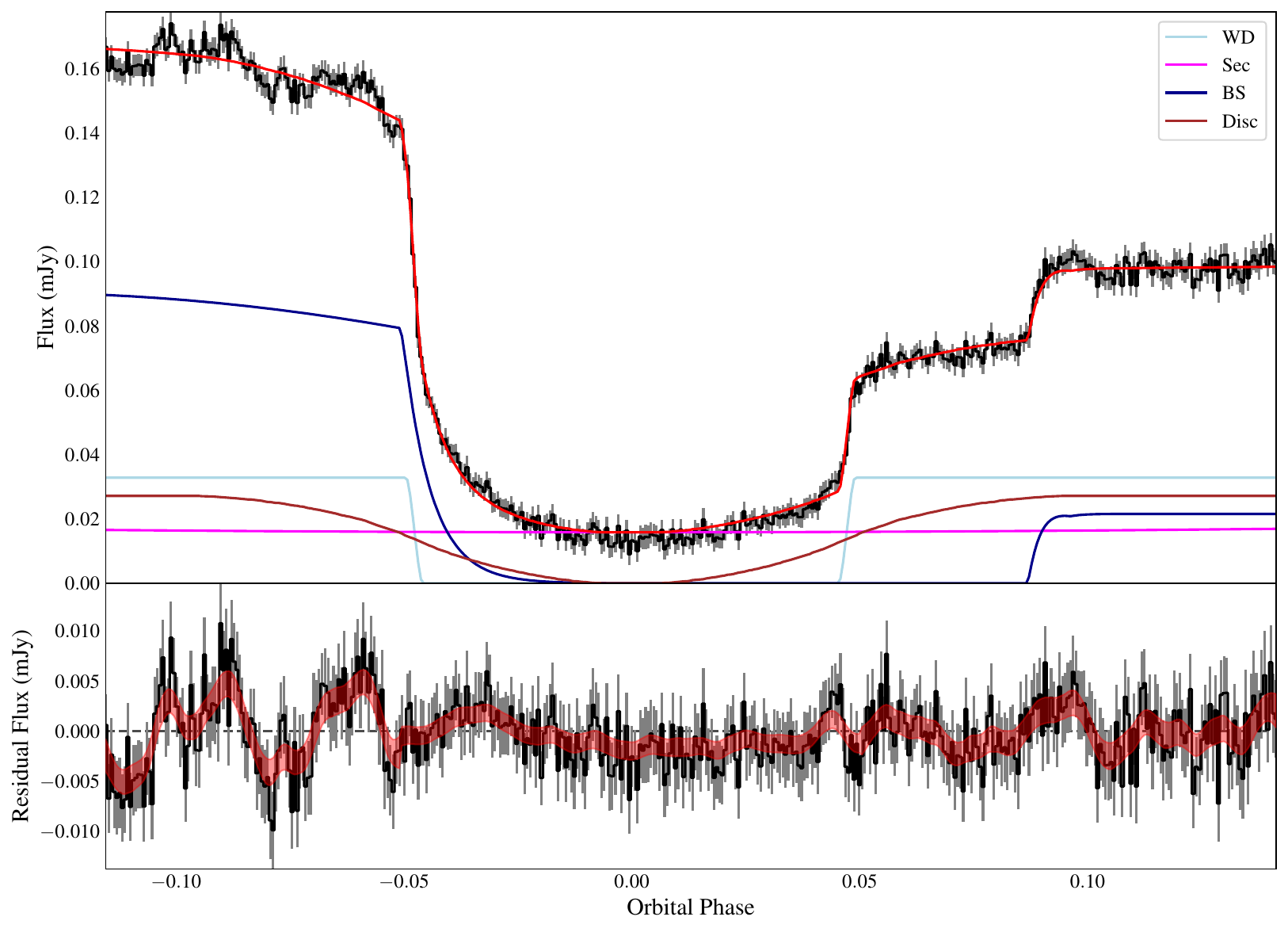}
\end{subfigure}
\begin{subfigure}[b]{.58\linewidth}
\includegraphics[width=\linewidth]{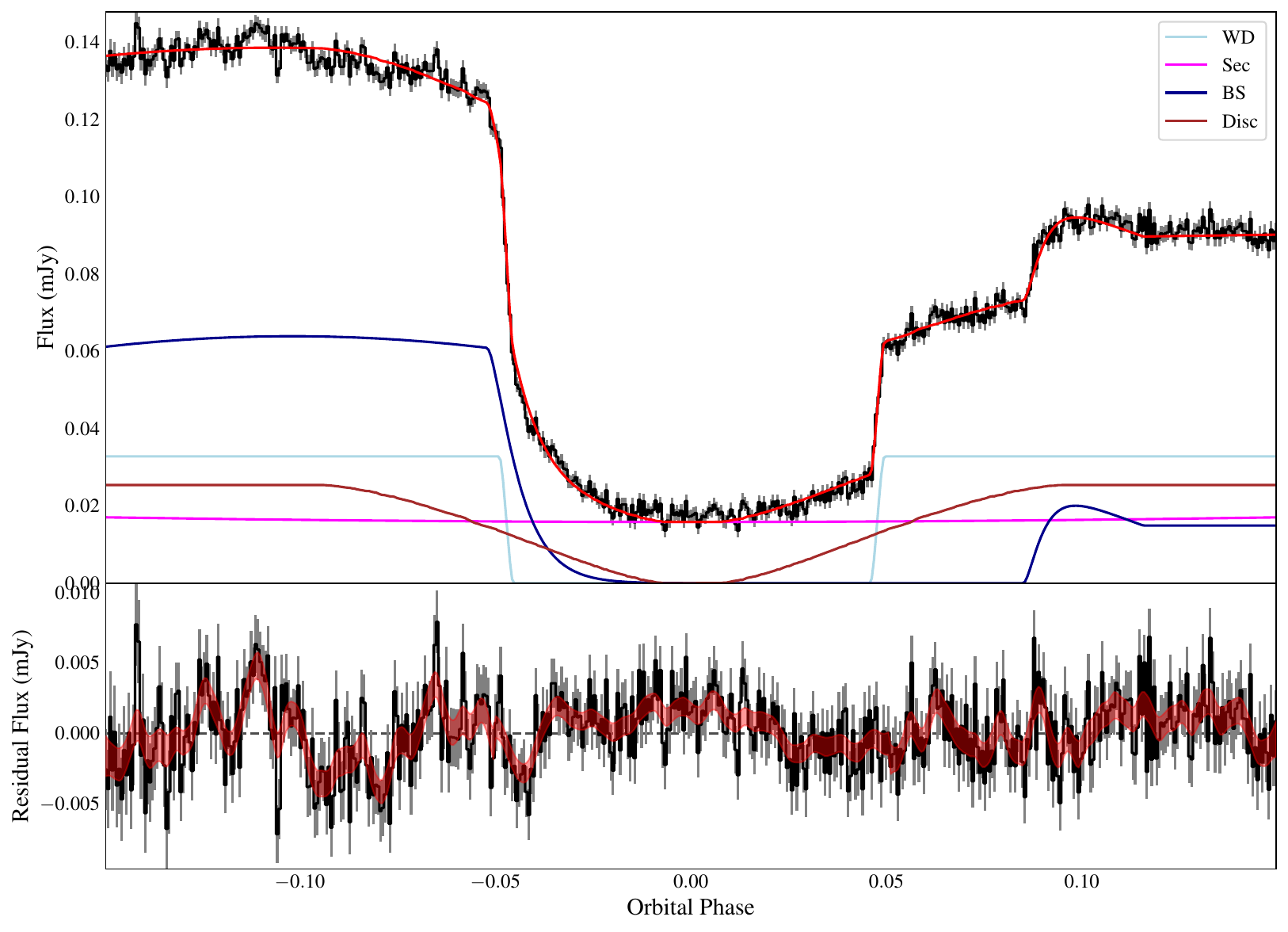}
\end{subfigure}
\caption{Phase folded and binned $g'$-band lightcurves together with model fits. From top to bottom, the panels show the 2016 data, the 2018-01-17 data, and the binned lightcurves of remaining data. Markers and colours are as described in figure~\ref{fig:rmodelfits}.}
\label{fig:gmodelfits}
\end{figure*}

\begin{figure}
\centering
\begin{subfigure}[b]{.99\linewidth}
\includegraphics[width=\linewidth]{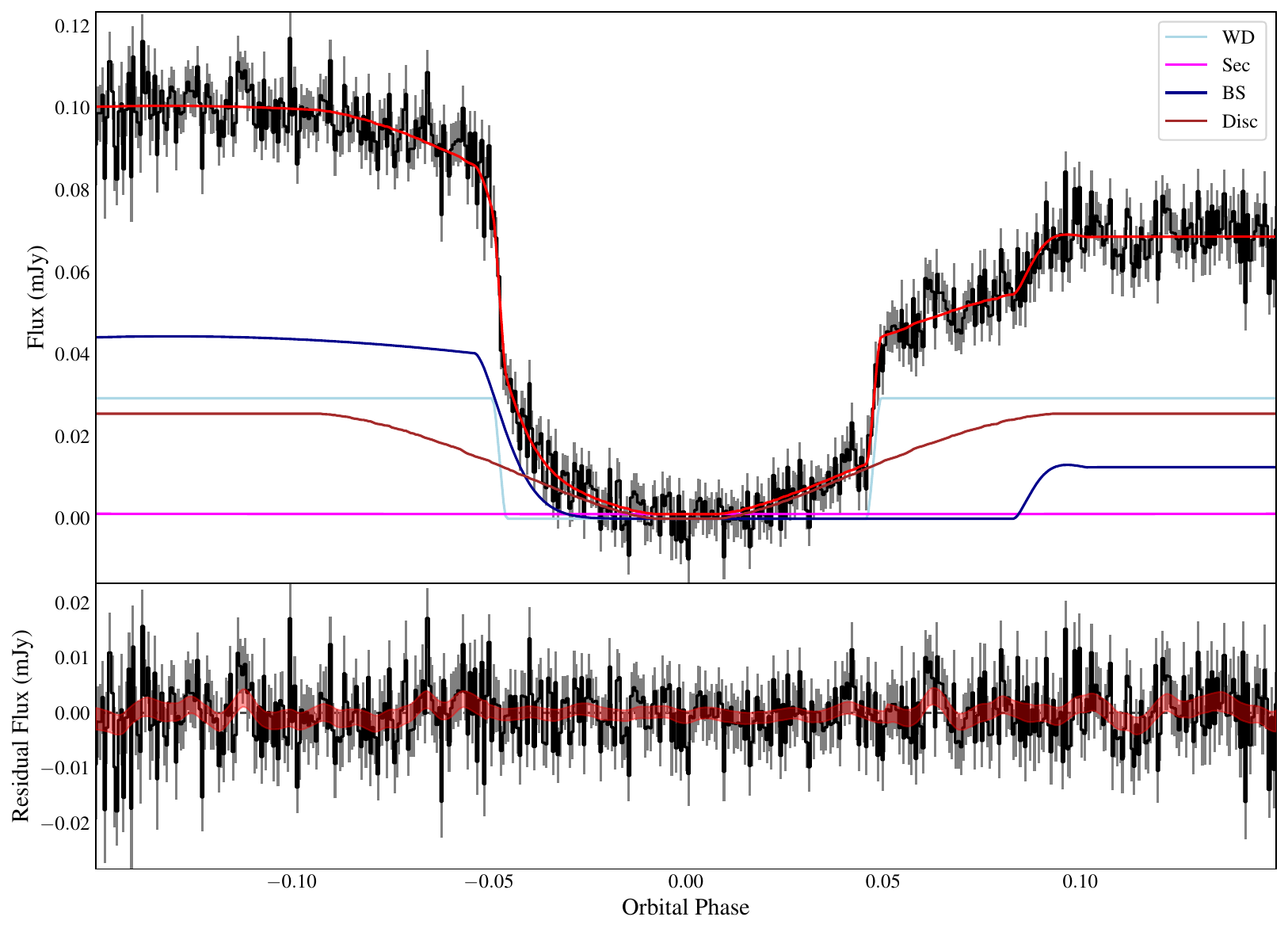}
\end{subfigure}
\caption{Phase folded and binned $u'$-band lightcurve together with model fits. Markers and colours are as described in figure~\ref{fig:rmodelfits}.}
\label{fig:umodelfits}
\end{figure}

With improved confidence that accurate system parameters can be retrieved from eclipses with blended ingresses, we proceed to model the CSS131106 eclipses. For the $g$- and $r$-band lightcurves, we produced three binned lightcurves. The 2016 data were taken in a different filter set and cannot be binned together with the later data. The eclipse obtained on the night starting 2018-01-17 has evidence for an increased bright spot flux and so was also modelled separately. The remaining eclipses were combined to produce the third binned lightcurve.
These lightcurves, together with model fits described below, are shown in figures~\ref{fig:rmodelfits} and \ref{fig:gmodelfits} for the $r$- and $g$-bands respectively. For the $u$-band, the signal-to-noise ratios of the data in 2016 and the night starting 2018-01-17 were too low to produce useful lightcurves, so only a binned lightcurve of all the remaining $u$-band data was produced, shown in figure~\ref{fig:umodelfits}.

The lightcurves were modelled using the model and procedures described in \cite{wild22}. The mass ratio $q$, white dwarf eclipse width $\Delta \phi_{1/2}$, and scaled white dwarf radius $R_w/a$ were shared between all eclipse lightcurves. separate white dwarf, disc, bright spot and donor fluxes were used in each band (with separate parameters for the $g'$ and $g_s$ bands, for example). The remaining model parameters (see \citealt{wild22} for details) were allowed to vary for each eclipse. In total, the model has 105 free parameters. After an initial model fit to estimate the white dwarf effective temperature, we placed Gaussian priors on the white dwarf limb-darkening parameters, interpolated from the tables of \cite{gianninas13}.

To help convergence of the Markov Chain Monte Carlo (MCMC) fitting, we used the ensemble sampler implemented in the {\sc emcee} package \citep{foreman-mackey2013}, with 450 walkers and 45~000 steps, discarding the first 10~000 steps as burn-in. After fitting convergence is checked using the Gelman-Rubin statistic \citep{gelman92}, ensuring that all parameters have $\hat{R} < 1.01$. The MCMC chains are thinned by a factor of 10, which is larger than the autocorrelation time for all parameters. The best-fit parameters were taken as the median of the posterior distributions, with uncertainties estimated from the 16th and 84th percentiles.

\begin{figure}
\includegraphics[width=\linewidth]{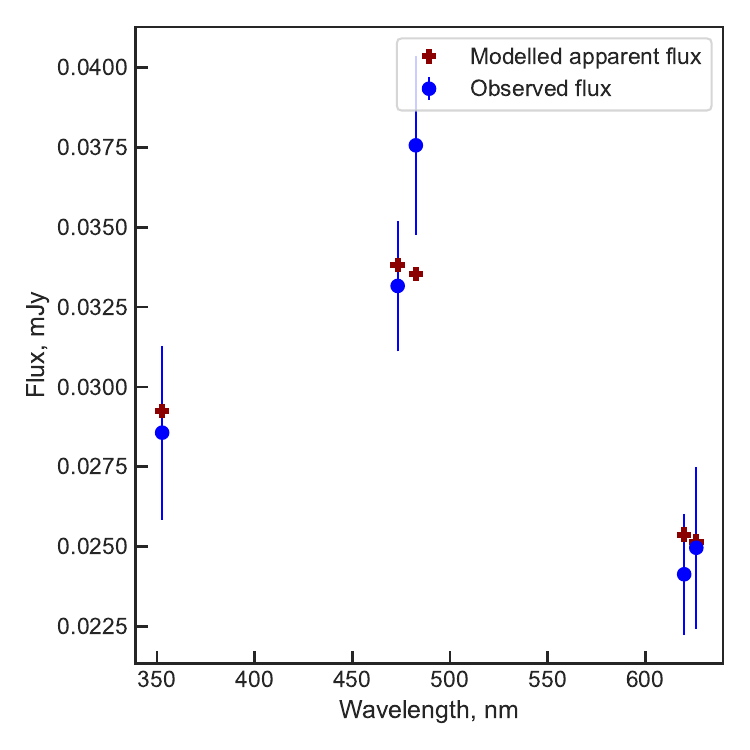}
\caption{Fit to white dwarf fluxes provided by the lightcurve modelling. The two data points for each of the SDSS $g'$ and $r'$ bands are from the 2016 and 2017/2018 data, which were taken in different filter sets and therefore have slightly different effective wavelengths. All measured fluxes are within 1.5$\sigma$ of the best fit model.}
\label{fig:fluxplot}
\end{figure}

\begin{figure*}
\includegraphics[width=\linewidth]{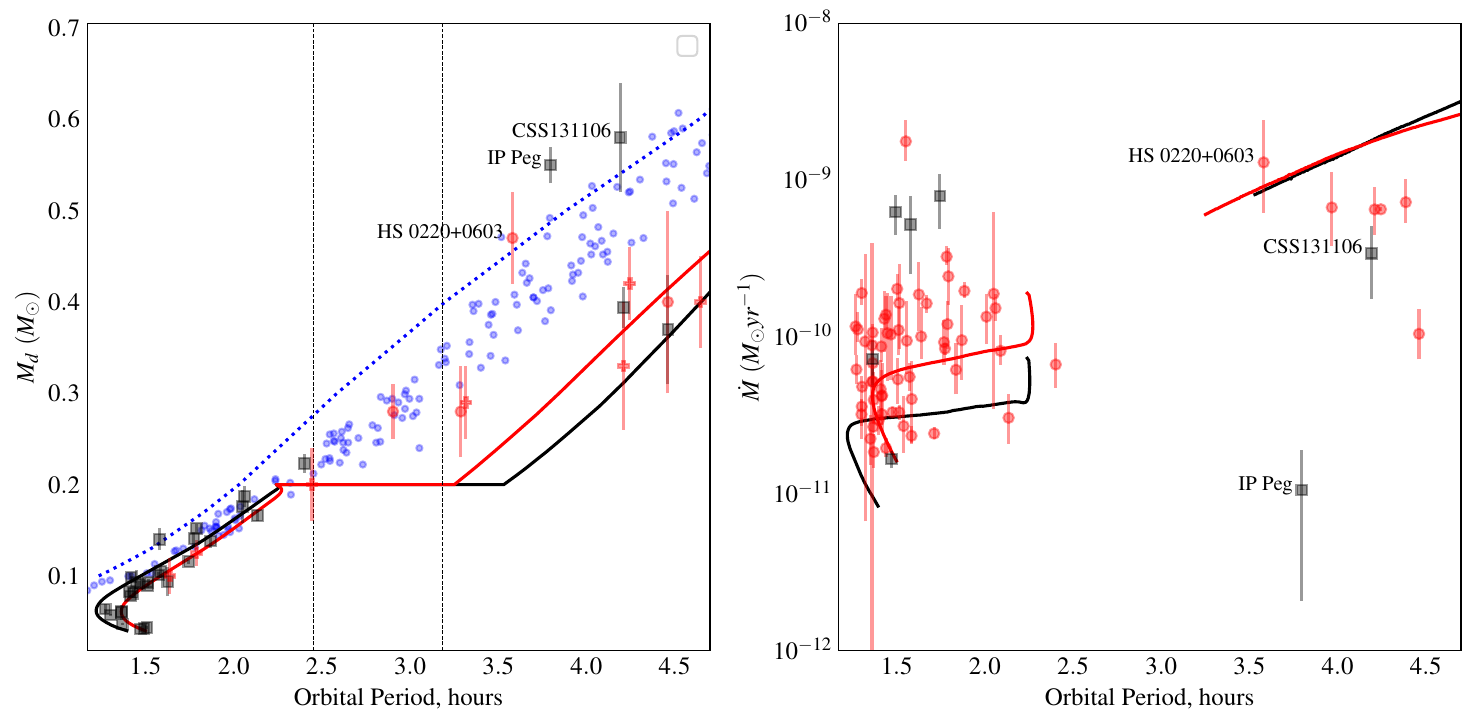}
\caption{Comparison between theoretical models and observational data for CVs. Left panel shows donor masses against orbital period, whilst the right panel shows accretion rate, derived from white dwarf masses and effective temperatures. Donor masses are compiled from \protect\cite{mcallister2019} and \protect\cite{wild22}. Donor masses measured using the eclipse method are shown as black points with error bars. In the left panel, red circles denote donor masses derived using the eclipse mapping method without correcting the white dwarf mass-radius relation used for the white dwarf effective temperature, and red crosses denote donor masses derived using radial velocities. In the right panel, black points with error bars are taken from eclipse mapping measurements \protect\citep{mcallister2019, wild22} whilst red circles show measurements from white dwarf spectroscopy \protect\citep{pala2022}. Where an effective temperature has been measured from both methods, we prefer the estimate from white dwarf spectroscopy. In both panels, evolutionary models from \protect\cite{knigge11b} are shown as solid lines. The black line denotes the ``standard'' track and the red line the ``optimal'' track, where the magnetic braking above the period gap is reduced by a factor of 0.66. Dashed vertical lines denote the limits of the period gap from \protect\cite{schreiber24}. The blue dashed line in the left-hand panel shows the period at which a main-sequence star would fill the Roche lobe using the \protect\cite{baraffe15} models. Blue points show the period at which a main-sequence star would fill its Roche lobe, using the observed M-dwarf sample from \protect\cite{mann15} with masses and radii measured using eclipsing binaries and interferometry.}
\label{fig:modelcomparison}
\end{figure*}

\subsection{System parameters}
\label{subsec:sys_params}
The model fits described in section~\ref{subsec:ecl_model} yield estimates of the mass ratio,  white dwarf eclipse width and scaled white dwarf radius. In addition they provide estimates of the white dwarf flux in each band. Following the procedures outlined in \cite{wild22}, we fit the white dwarf fluxes to the white dwarf cooling tracks of \cite{bergeron95}, with color corrections to the appropriate ULTRACAM filter, in order to estimate the white dwarf effective temperature $T_{\rm eff}$ and the parallax of the CV. The {\sc dustmaps} package was used to place a prior on the extinction using the Bayestar19 model \citep{green18, bayestar19}. A Gaussian prior is placed on the parallax using the GaiaDR3 measurement \citep{gaiadr3}. The fit to the white dwarf fluxes is shown in figure~\ref{fig:fluxplot}.

The model parameters from the lightcurve fits can be converted into physical system parameters by adopting a white dwarf mass-radius relationship. By assuming a white dwarf mass, $M_w$, and using the mass ratio from lightcurve fitting we can estimate the total mass. Combined with the orbital period and Kepler's third law this provides the orbital separation $a$. If the assumed white dwarf mass is correct, the radius from the mass-radius relationship should match the scaled radius from lightcurve fitting, $R_w/a$, allowing us to solve for the white dwarf mass. We repeat this process for every MCMC step from lightcurve fitting to construct posterior probability distributions on the system parameters.

We used the same procedure as described in \cite{wild22}, using the mass-radius relationships of \cite{Wood_1995} for carbon-oxygen core white dwarfs with thick hydrogen layers. The mass-radius relationship is corrected for temperature using the temperature estimates derived from the white dwarf fluxes. The resulting system parameters are presented in table~\ref{tab:sys_params}.

\begin{table}
\centering
\caption{System parameters for CSS131106.}
\label{tab:sys_params}
\begin{tabular}{lcc}
\hline
Parameter & Median & Uncertainty \\
\hline
\multicolumn{3}{c}{Lightcurve Fitting Parameters} \\
\hline
Mass Ratio $q$ & 0.81 & 0.06 \\
inclination $i$ (degrees) & 78.5 & 0.7 \\
$R_w/a$ & 0.015 & 0.001\\
\hline
\multicolumn{3}{c}{Priors} \\
\hline
Parallax (mas) & 1.30 & 0.10 \\
E(B-V) & 0.055 & 0.005 \\
\hline
\multicolumn{3}{c}{Parameters from WD fluxes} \\
\hline
$T_{\rm eff}$ (K) & 18~500 & 2~000 \\
Parallax (mas) & 1.30 & 0.10 \\
\hline
\multicolumn{3}{c}{Derived Parameters} \\
\hline
$M_{w}$ (\msun) & 0.72 & 0.04 \\
$R_{w}$ ($R_{\odot}$) & 0.011 & 0.005 \\
$\log(g_w [cgs])$ & 8.18 & 0.06 \\
$M_{d}$ (\msun) & 0.58 & 0.06 \\
$R_{d}$ ($R_{\odot}$) & 0.52 & 0.02 \\
$a$ ($R_{\odot}$) & 1.43 & 0.03 \\
$K_d$ (km s$^{-1}$) & 182 & 10 \\
$K_w$ (km s$^{-1}$) & 79 & 1 \\
$\dot{M} $ ($M_{\odot}$ yr$^{-1}$) & $3 \times 10^{-10}$ & $1 \times 10^{-10}$ \\
\end{tabular}
\end{table}

\section{Discussion}
\label{sec:discussion}
As noted by \cite{knigge11b}, the donor masses and white dwarf temperatures in CVs are valuable probes of the mass transfer rates and angular momentum loss mechanisms in CVs. The white dwarf temperature is a tracer of the long term mass loss $\dot{M}$, since it is produced by the compression of the white dwarf by the accreted matter, rather than the accretion heating itself \citep{Townsley_Bildsten_2003,Townsley_Gansicke_2009}. For the white dwarfs in CVs, the white dwarf temperature traces the mass transfer rate average over $\tau_{w} \sim 10^4$ years \citep{Townsley_Bildsten_2003, knigge11b}. \cite{pala2022} provide a compilation of white dwarf masses and effective temperatures from ultraviolet spectroscopy. The white dwarf masses and effective temperatures constrain $\dot{M}$ via:
\begin{equation}
L_{w} = 4 \pi R_{w}^2 \sigma T_{\rm eff}^4 = 6 \times 10^{-3} L_{\odot} \left(\frac{\dot{M}}{10^{-10} M_{\odot} yr^{-1}}\right) \left(\frac{M_{w}}{0.9 M_{\odot}}\right)^{0.4}.
\end{equation}
Accretion rates measured this way are shown in the right-hand panel of figure~\ref{fig:modelcomparison}. In addition, we also provide estimates of accretion rates using white dwarf masses and effective temperatures from the photometric eclipse method, although the direct measures of \cite{pala2022} are preferred where available.

Donor masses are a valuable complementary probe of mass loss rates as they trace the mass loss rate over a timescale $\tau_{d} \approx 0.05 \, \tau_{\rm kh}$ years, where $\tau_{\rm kh}$ is the Kelvin-Helmholtz timescale of the donor \citep{knigge11b}. This is significantly longer than the timescale traced by the white dwarf temperature, and so donor masses can be used to see if the mass loss rates have been stable over long timescales. However, since $\tau_{d}$ can be comparable to the evolutionary timescale of the CV, the donor mass may be affected by the mass loss {\em history}, which can complicate comparisons (see the right-panel of Figure~1 in \citealt{knigge11b}).
A compilation of donor masses in CVs are shown in the left-hand panel of figure~\ref{fig:modelcomparison}.

Looking at the donor mass first, CSS131106 joins two other CVs which have very large donor masses for their orbital period: IP Peg \citep{copperwheat10} and HS~0220+0603 \citep{rodriguez-gil15}. Since the donor fills the Roche Lobe at a shorter orbital period than expected for its mass, the donor star must be smaller than expected. Indeed, figure~\ref{fig:modelcomparison} shows that the donor radius in CSS131106 is consistent with theoretical radii for a main-sequence M-dwarf. Whilst M-dwarfs tend to be slightly larger than predicted by models \cite[e.g.][]{mann15, brown22}, there is considerable scatter in the observed radii \citep{parsons18}, and some M-dwarfs in detached binaries have radii consistent with model predictions. Figure~\ref{fig:modelcomparison} shows that the donors in IP~Peg, HS~0220+0603 and CSS131106 are quite small, even compared to detached M-dwarfs of similar mass. The donor masses in these three CVs would therefore suggest that, averaged over $\tau_d$, the mass loss rate $\dot{M}$ in these systems is low enough to allow the donor to remain in thermal equilibrium.

The white dwarf temperatures in the three CVs show that, averaged over the last few thousand years, the three systems have experienced quite different mass loss rates. For IP~Peg, the inferred $\dot{M}$ is very low. However, HS~0220+0603 has a $\dot{M}$ that is entirely consistent with the ``optimal'' evolutionary track of \cite{knigge11b}. The present-day accretion rate in HS~0220+0603 is also high; the system is a nova-like variable of SW Sex type, which show bright accretion discs and high accretion luminosities \citep{rodriguez-gil07}. CSS131106 has $\dot{M} = 3 \pm 1 \times 10^{-10}$~\msun{}~yr$^{-1}$. This value is somewhat lower than HS~0220+0603, but it is consistent with other dwarf novae at this orbital period \citep{pala2022}. 

Whilst the donor masses and white dwarf temperatures may appear to be in tension, this is not in fact the case. This is because the large donor masses imply relatively short Kelvin-Helmholtz timescales of $\tau_{\rm kh} \approx 5 \times 10^8$ years ($\tau_d \approx 3\times10^7$ years). As a result, the donors can remain close to thermal equilibrium even with relatively high mass loss rates. By calculating evolutionary tracks with MESA \citep{paxton11, paxton13, paxton15, paxton18, paxton19,jermyn23} of stars steadily losing mass so that they reach a mass of $\sim 0.55$~\msun{} at an age of 2 Gyr, we find that mass loss rates of up to $\sim 5 \times 10^{-10}$~\msun{}~yr$^{-1}$ can be sustained without significantly inflating these donor stars above their thermal equilibrium radius. Therefore, for CSS131106 and IP~Peg, the mass loss rates inferred from the white dwarf temperatures are not in conflict with the small donor radii. However, for HS~0220+0603, the inferred mass loss rate of $\sim 10^{-9}$~\msun{}~yr$^{-1}$ appears too high to allow the donor to remain close to thermal equilibrium. Therefore, it is likely that HS~0220+0603 is undergoing a short-term episode of high mass transfer, and that its long-term average mass loss rate is lower.

However, the high donor masses still require an explanation. If the mass loss rates inferred from the white dwarf temperatures are representative of the long-term mass loss rate, these stars would rapidly lose mass, increasing the thermal timescale and driving them from thermal equilibrium. Hence once mass transfer begins, these systems would converge onto the main evolutionary track for CVs over a timescale of roughly $\tau_d \sim 10^7$ years. Therefore, if the mass transfer rates from the white dwarf temperatures reflect the long term average, these systems must have begun mass transfer within the last $\sim 10^7$ years. This seems unlikely a-priori.

Alternatively, it may be the case that these systems have very low long-term average mass loss rates, allowing them to stay near thermal equilibrium. Models of CV evolution predict mass loss rates of $\sim 10^{-9}$~\msun{}~yr$^{-1}$ at these orbital periods \citep{knigge11b}, and we have shown that mass loss rates of a few times $10^{-10}$~\msun{}~yr$^{-1}$ can be sustained without significantly inflating the donors. The donor masses in these systems can then be explained if the angular momentum loss rates are $\sim 10$ times lower than in most CVs, or if the systems spend a large fraction of their time ($\sim$\,90\%) in a detached state with very low mass loss rates. One plausible mechanism for this is hibernation following a nova eruption \citep{shara86}. As discussed in \cite{shara86}, the detached phase is expected to last between $10^3$--$10^6$ years, with the smaller of those values more appropriate for long period CVs. The recurrence timescale for novae in long period CVs is $10^4$--$10^5$ years \cite{knigge11a}, so the duty cycle of hibernation may be sufficient to explain the low average mass transfer rates required to keep the donors close to thermal equilibrium. However, the hibernation scenario does not explain why only these three systems have such low donor radii; if hibernation is common, we would expect to see many more CVs with small donors. Our conclusion is therefore that these systems must have formed very recently, or be experiencing unusually low angular momentum loss rates compared to most CVs.

A third alternative is that the donor stars in these systems are not typical main sequence stars. We discuss this possibility further below. As discussed in \cite{knigge11b}, the $M$--$T_{\rm eff}$ relation for stars with a large convective envelope is insensitive to mass loss, so we expect CV donors to have similar effective temperatures as main-sequence stars of the same mass \citep[see][for detailed calculations]{baraffe00}. The donor stars in HS~0220+0603 and IP~Peg have spectral types of M5.5V \citep{rodriguez-gil15} and M4V \citep{littlefair01} respectively. Main sequence stars with similar masses have spectral types of M0--M1V. Spectral types of M4--M5V imply masses of 0.15--0.2~\msun{} \citep{kirkpatrick24}, roughly half the measured masses of the donors in these two systems. No spectra of the donor star in CSS131106 exist, but we can obtain a crude estimate from the contribution of the donor star estimated from the lightcurve modelling. This gives a colour for the donor star of $g'-r' = 1.50 \pm 0.04$, consistent with a spectral type of M4--M5V \citep{Covey2007}. The donor stars in all three systems appear significantly cooler than expected for their masses, and spectroscopy of the donor star in CSS131106 would be of interest.

We are unable to identify a plausible mechanism that can explain both the low effective temperatures and the small radii of these donor stars. The donor stars in these three systems are small even when compared to detached main-sequence stars of the same mass. Starspots could lower the effective temperature at a given mass. Converting the observed spectral types into effective temperatures using the scale presented in \cite{pecault13}, we find values of $T_{\rm eff} \sim 3100$~K, whereas based on the masses we would expect $T_{\rm eff} \sim 3700$~K. The SPOTS models of \cite{somers20} show that, for $M=0.5$~\msun, extensive starspot coverage could reduce the effective temperature by $\sim 500$~K. However, such large starspot coverages also inflate the stellar radius by 10--15\%. Low metallicity would make the donor stars smaller, but would substantially increase the effective temperature \cite[e.g][]{rebassa-mansergas19}, contrary to what is observed. Therefore the small radii and cool temperatures of the donors in these three CVs remains an unsolved problem.

\section{Conclusions}
We demonstrate that accurate system parameters can be retrieved from the eclipse lightcurves of CVs, even when the white dwarf and bright spot ingresses are blended. We apply the photometric eclipse method to measure the component masses and white dwarf temperature in the long period CV CSS131106 J052412+004148, finding a mass ratio of $q = 0.81 \pm 0.06$ and a donor mass of $M_d = 0.58 \pm 0.06$~\msun{}. The white dwarf has a mass of $M_w = 0.72 \pm 0.04$~\msun{} and an effective temperature of $T_{\rm eff} = 18~500 \pm 2~000$~K, implying a mass transfer rate of $\dot{M} = 3 \pm 1 \times 10^{-10}$~\msun{}~yr$^{-1}$. The donor star is significantly smaller than expected for its mass, or - equivalently - more massive than expected for its orbital period.  CSS131106 J052412+004148 joins two other long period CVs (IP~Peg and HS~0220+0603) with similarly small donors. Puzzlingly, the donor stars in all three systems appear significantly cooler than expected for their masses.

The unusually small donors in these three systems can be explained if they have only recently begun mass transfer, or if they are experiencing unusually low long-term mass transfer rates compared to most CVs. However, the low effective temperatures of the donor stars remain unexplained.

\section*{Acknowledgements}
SPL acknowledges support from the Science and Technology Facilities Council (grant ST/Z000033/1).
SGP acknowledges support by the Science and Technology Facilities Council (grant ST/B001174/1). This project has received funding from the European Research Council under the European Union's Horizon 2020 research and innovation programme (Grant agreement numbers 101002408 - MOS100PC).
IP acknowledges support from the Royal Society through a University Research Fellowship (URF\textbackslash R1\textbackslash 231496).
\section*{Data Availability}

The data underlying this article will be shared on reasonable request
to the corresponding author.



\bibliographystyle{mnras}
\bibliography{new_refs, refs, refs2, refs3, refs4} 








\bsp	
\label{lastpage}
\end{document}